\documentclass[12pt]{article}
\usepackage{nature,citesupernumber,citecollapse}
\usepackage{graphicx,color}
\usepackage[english]{babel}
\makeatletter
\renewcommand\@biblabel[1]{#1.}

\def\@cite#1#2{$^{\mbox{\scriptsize #1\if@tempswa , #2\fi}}$}

\newcommand{\spacing}[1]{\renewcommand{\baselinestretch}{#1}\large\normalsize}
\spacing{2}

\def\@maketitle{%
  \newpage\spacing{1}\setlength{\parskip}{12pt}%
    {\Large\bfseries\noindent\sloppy \textsf{\@title} \par}%
    {\noindent\sloppy \@author}%
}

\renewenvironment{abstract}{%
    \setlength{\parindent}{0in}%
    \setlength{\parskip}{0in}%
    \bfseries%
    }{\par\vspace{-6pt}}

\newenvironment{addendum}{%
    \setlength{\parindent}{0in}%
    \small%
    \begin{list}{Acknowledgements}{%
        \setlength{\leftmargin}{0in}%
        \setlength{\listparindent}{0in}%
        \setlength{\labelsep}{0em}%
        \setlength{\labelwidth}{0in}%
        \setlength{\itemsep}{12pt}%
        }
    }
    {\end{list}\normalsize}

\makeatother

\def\Rstar{\hbox{$R_{\star}$}}              

\def\irc{IRC\,+10216}

\newcount\longrefs
\def\aap{\ifnum\longrefs=1 {Astron.\ Astrophys.}\else 
                           {A\hbox{\rm \&}A}\fi}
\def\aapl{\ifnum\longrefs=1 {Astron.\ Astrophys.\ Lett.}\else 
                           {A\hbox{\rm \&}A}\fi}
\def\aapr{\ifnum\longrefs=1 {Astron.\ Astrophys.\ Rev.}\else 
                            {A\hbox{\rm \&}AR}\fi}
\def\aaps{\ifnum\longrefs=1 {Astron.\ Astrophys.\ Suppl.}\else 
                            {A\hbox{\rm \&}AS}\fi}
\def\aj{\ifnum\longrefs=1 {Astron.\ J.}\else 
                          {AJ}\fi} 
\def\ao{\ifnum\longrefs=1 {Applied Optics}\else 
                           {Appl.\ Opt.}\fi} 
\def\aspcs{\ifnum\longrefs=1 {Astron.\ Soc.\ Pacific Conf. Series}\else 
                           {ASP Conf.\ Ser.}\fi} 
\def\apj{\ifnum\longrefs=1 {Astrophys.\ J.}\else 
                           {ApJ}\fi} 
\def\apjl{\ifnum\longrefs=1 {Astrophys.\ J.\ Lett.}\else 
                            {ApJ}\fi} 
\def\aplett{\ifnum\longrefs=1 {Astrophys.\ J.\ Lett.}\else 
                            {ApJ}\fi} 
\def\apjs{\ifnum\longrefs=1 {Astrophys.\ J.\ Suppl.}\else 
                            {ApJS}\fi}
\def\apss{\ifnum\longrefs=1 {Astrophys.\ and Space Science}\else 
                            {Ap\hbox{\rm \&}SS}\fi}
\def\araa{\ifnum\longrefs=1 {Ann.\ Rev.\ Astron.\ Astrophys.}\else 
                            {ARA\hbox{\rm \&}A}\fi}
\def\azh{\ifnum\longrefs=1 {Astronomicheskii Zhurnal}\else 
                            {Astron.\ Zhur.}\fi}
\def\baas{\ifnum\longrefs=1 {Bull.\ Am.\ Astron.\ Soc.}\else 
                            {BAAS}\fi}
\def\bain{\ifnum\longrefs=1 {Bull.\ Astronom.\ Institutes Netherlands}\else
                            {Bull.\ Astr.\ Inst.\ Neth.}\fi}
\def\gca{\ifnum\longrefs=1 {Geochim.\ Cosmochim.\ Acta}\else 
                           {Geochim.\ Cosmochim.\ Acta}\fi}
\def\grl{\ifnum\longrefs=1 {Geophys.\ Res.\ Lett.}\else 
                           {Geoph.\ Res.\ Lett.}\fi}
\def\iaucirc{\ifnum\longrefs=1 {IAU Circulars}\else 
                          {IAU Circ.}\fi}
\def\ip{\ifnum\longrefs=1 {in press}\else 
                          {in press}\fi}
\def\jchemp{\ifnum\longrefs=1 {J.\ Chem.\ Phys.}\else 
                           {J.\ Chem.\ Phys.}\fi}  
\def\jcp{\ifnum\longrefs=1 {J.\ Chem.\ Phys.}\else 
                           {J.\ Chem.\ Phys.}\fi}  
\def\jgr{\ifnum\longrefs=1 {J.\ Geophys.\ Res.}\else 
                           {J.\ Geophys.\ Res.}\fi}  
\def\jmolspec{\ifnum\longrefs=1 {J.\ Mol.\ Spectrosc.}\else 
                           {J.\ Mol.\ Spectrosc.}\fi}  
\def\jqsrt{\ifnum\longrefs=1 {J.\ Quant.\ Spectrosc.\ Radiat.\ Transfer}\else 
                           {J.\ Quant.\ Spectrosc.\ Radiat.\ Transfer}\fi}  
\def\jrasc{\ifnum\longrefs=1 {J.\ Royal Astron.\ Soc.\ Canada}\else 
                           {JRAS Can.}\fi}  
\def\mnras{\ifnum\longrefs=1 {Mon.\ Not.\ Roy.\ Astron.\ Soc.}\else 
                             {MNRAS}\fi} 
\def\nat{\ifnum\longrefs=1 {Nature}\else 
                           {Nat}\fi}
\def\pasj{\ifnum\longrefs=1 {Pub.\ Astron.\ Soc.\ Japan}\else 
                            {PASJ}\fi} 
\def\pasp{\ifnum\longrefs=1 {Pub.\ Astron.\ Soc.\ Pacific}\else 
                            {PASP}\fi} 
\def\physscr{\ifnum\longrefs=1 {Physica Scripta}\else 
                            {Phys.\ Scrip.}\fi} 
\def\planss{\ifnum\longrefs=1 {Planetary \& Space Science}\else 
                            {Plan. \& Space Sci.}\fi} 
\def\procspie{\ifnum\longrefs=1 {Proc.\ SPIE}\else 
                            {Proc.\ SPIE}\fi} 
\def\qjras{\ifnum\longrefs=1 {Quarterly J.\ Royal Astron.\ Soc.}\else 
                            {QJRAS}\fi} 
\def\sa{\ifnum\longrefs=1 {Soviet Astron..}\else 
                               {Sov.\ Astron.}\fi}
\def\skytel{\ifnum\longrefs=1 {Sky \& Telescope}\else 
                            {Sky \& Tel.}\fi} 
\def\solphys{\ifnum\longrefs=1 {Solar Phys.}\else 
                               {Solar Phys.}\fi}
\def\ssr{\ifnum\longrefs=1 {Space Science Rev.}\else 
                               {Space\ Sci.\ Rev.}\fi}

\title{Warm water vapour in the sooty outflow from a luminous carbon star}

\author{L. Decin$^{1,2}$,
	M. Ag\'undez$^{3,7}$,
        M.J. Barlow$^{4}$,
 	F. Daniel$^{3}$,
 	J. Cernicharo$^{3}$,
	R. Lombaert$^1$,
	E. De Beck$^1$,
        P. Royer$^{1}$,          
        B. Vandenbussche$^{1}$,
         R. Wesson$^{4}$,
         E.T. Polehampton$^{5,6}$,
         J.A.D.L. Blommaert$^{1}$,
          W. De Meester$^{1}$,
         K. Exter$^{1}$,
         H. Feuchtgruber$^{8}$,
         W.K. Gear$^{9}$,
         H.L. Gomez$^{9}$,
         M.A.T. Groenewegen$^{10}$,
	M. Gu\'elin$^{16}$,
         P.C. Hargrave$^{9}$,
         R. Huygen$^{1}$,
         P. Imhof$^{11}$,
         R.J. Ivison$^{12}$,
         C. Jean$^{1}$,
	C. Kahane$^{17}$,
	F. Kerschbaum$^{14}$,
         S.J. Leeks$^{5}$,
         T. Lim$^{5}$,
	M. Matsuura$^{4,15}$,
         G. Olofsson$^{13}$,
	T. Posch$^{14}$,
         S. Regibo$^{1}$,
         G. Savini$^{4}$,
         B. Sibthorpe$^{12}$,
         B.M. Swinyard$^{5}$,
 	J.A. Yates$^{4}$,
	\&
        C. Waelkens$^{1}$
 }

\begin{document}

\maketitle

\newenvironment{affiliations}{%
    \setcounter{enumi}{1}%
    \setlength{\parindent}{0in}%
    \slshape\sloppy%
    \begin{list}{\upshape$^{\arabic{enumi}}$}{%
        \usecounter{enumi}%
        \setlength{\leftmargin}{0in}%
        \setlength{\topsep}{0in}%
        \setlength{\labelsep}{0in}%
        \setlength{\labelwidth}{0in}%
        \setlength{\listparindent}{0in}%
        \setlength{\itemsep}{0ex}%
        \setlength{\parsep}{0in}%
        }
    }{\end{list}\par\vspace{12pt}}

\small{
\begin{affiliations}
	\item Instituut voor Sterrenkunde, Katholieke Universiteit Leuven, Celestijnenlaan 200D, 3001 Leuven, Belgium
 		\item
	Sterrenkundig Instituut Anton Pannekoek, University of Amsterdam, Science Park 904, NL-1098 Amsterdam, The Netherlands 
	\item
	   Laboratory of Molecular Astrophysics, Department of Astrophysics, CAB, INTA-CSIC, Ctra de Ajalvir, km 4, 28850 Torrej\'on de Ardoz, Madrid Spain 
       \item
             Dept of Physics \& Astronomy, University College London, Gower St, London WC1E 6BT, UK
       \item
             Space Science and Technology Department, Rutherford Appleton Laboratory, Oxfordshire, OX11 0QX, UK
       \item
            Institute for Space Imaging Science, University of Lethbridge, Lethbridge, Alberta, T1J 1B1, Canada
    \item
    LUTH, Observatoire de Paris-Meudon, 5 Place Jules Janssen, 92190 Meudon, France 
    \item
             Max-Planck-Institut f\"ur extraterrestrische Physik, Giessenbachstrasse, 85748 Garching, Germany
        \item
             School of Physics and Astronomy, Cardiff University, Queens Buildings, The Parade, Cardiff, CF24 3AA, UK
    \item
    Royal Observatory of Belgium, Ringlaan 3, B-1180 Brussels, Belgium 
        \item
            Blue Sky Spectroscopy, 9/740 4 Ave S, Lethbridge, Alberta T1J 0N9, Canada 
    \item
    UK Astronomy Technology Centre, Royal Observatory Edinburgh, Blackford Hill, Edinburgh EH9 3HJ, UK 
         \item
             Dept of Astronomy, Stockholm University, AlbaNova University Center, Roslagstullsbacken 21, 10691 Stockholm, Sweden
         \item
            University of Vienna, Department of Astronomy, T{\"u}rkenschanzstra\ss{}e 17, A-1180 Vienna, Austria 
    \item
    Mullard Space Science Laboratory, University College London,
Holmbury St. Mary, Dorking, Surrey RH5 6NT, United Kingdom 
\item
Institut de Radioastronomie Millim\'etrique, 300 rue de la Piscine, 38406 St. Martin d'H\'eres, France 
\item
Laboratoire d\'\,Astrophysique, Observatoire de Grenoble, BP 53, F-38041 Grenoble Cedex 9, France
\end{affiliations}
}

\begin{abstract}
In 2001, the discovery of circumstellar water
vapour around the ageing carbon star IRC\,+10216 was
announced\cite{Melnick2001Natur.412..160M}. This detection challenged the current understanding of chemistry in old stars, since water vapour was predicted to be absent in carbon-rich stars\cite{Willacy1998A&A...330..676W}. Several explanations for the occurrence of water 
vapour were postulated, including the vaporization of icy bodies (comets or dwarf planets)
in orbit around the star\cite{Melnick2001Natur.412..160M}, grain surface
reactions\cite{Willacy2004ApJ...600L..87W}, 
and photochemistry in the outer circumstellar
envelope\cite{Agundez2006ApJ...650..374A}. However, the only
water line detected so far from one carbon-rich evolved star can not
discriminate, by itself, between the different mechanisms
proposed. Here we report on the detection by the Herschel satellite\cite{Pilbratt2010} of dozens of water vapour lines
in the far-infrared and sub-millimetre spectrum of IRC\,+10216, including
some high-excitation lines with energies corresponding to
$\sim$1000\,K. The emission of these high-excitation water lines
can only be explained if water vapour is present in the warm inner
region of the envelope. A plausible explanation for
the formation of warm water vapour appears to be the penetration of 
ultraviolet (UV) photons  deep into a clumpy circumstellar
envelope. This mechanism triggers also the formation of other molecules such as ammonia, whose observed abundances\cite{2006ApJ...637..791H} are much higher than hitherto predicted\cite{Cherchneff1992ApJ...394..703C}.

\end{abstract}

For stars less massive than about nine times the mass of the sun,
the last major nuclear burning phase is as an Asymptotic Giant Branch
(AGB) star\cite{Iben1983ARA&A..21..271I}. A natural chemical division is created
between C (carbon-rich AGB) stars (with C/O ratio $>$1, and hence a surplus
of carbon to drive an organic chemistry in the envelope,
M-type AGB stars (with C/O ratio $<$1, yielding the formation of
oxygen-bearing molecules) and S-type AGB stars (with C/O ratio $\sim$1).
The detection of the ground-state transition of ortho-water, ortho-H$_2$O($1_{1,0}-1_{0,1}$), in the
envelope around the C-star IRC\,+10216 came as a surprise\cite{Melnick2001Natur.412..160M}, as in thermodynamic equilibrium (TE) chemistry no oxygen-rich molecules (except CO) are expected in a carbon-rich environment. The
vaporization of a collection of icy bodies (comets or dwarf planets)
in orbit around the star was invoked to explain the
presence of water vapour in this carbon-rich environment. It was predicted that water
should be released in the intermediate envelope at radii larger than
$(1-5)\times10^{15}$cm\emph{\cite{Melnick2001Natur.412..160M}}. Later on, two other distinct mechanisms were
considered as possible sources of water vapour observed in
IRC\,+10216, each one making a specific prediction for the H$_2$O
spatial distribution in the envelope: grain-surface reactions such as
Fischer-Tropsch catalysis on the surface of small
grains\cite{Willacy2004ApJ...600L..87W}, which would imply that water
reaches its maximum abundance at a distance around $2\times10^{15}$cm,
and formation in the outer envelope through the radiative association
of atomic oxygen and molecular
hydrogen\cite{Agundez2006ApJ...650..374A}. It has also been suggested
that water could be formed in the warm and dense inner
envelope\cite{GonzalesAlfonso2007ApJ...669..412G}, although no
specific formation mechanism has been proposed for such an origin. 

On 12 and 19 November 2009, IRC\,+10216 
was observed with the SPIRE\cite{Griffin2010}  and  PACS\cite{Poglitsch2010} 
spectrometers on board the Herschel Space
Observatory\cite{Pilbratt2010}. Spectroscopic observations were obtained
between 55 and 670\,$\mu$m, at spectral resolving powers between
300 and 4500\emph{\cite{Decin2010}}. Currently,  many different molecules and their
isotopologues have been identified: $^{12}$CO, $^{13}$CO, C$^{18}$O, H$^{12}$CN,
H$^{13}$CN, NH$_3$, SiS, SiO, CS, C$^{34}$S, $^{13}$CS, C$_3$,
C$_2$H, HCl,  H$^{37}$Cl, ortho-H$_2$O, and para-H$_2$O. The
detection of the low-excitation ortho-H$_2$O($2_{1,2}-1_{0,1}$)
transition  at 179.5\,$\mu$m was anticipated, since the energy
difference between the  $1_{1,0}$ and $2_{1,2}$ level is only
53\,K, but the discovery of high-excitation ortho-H$_2$O lines with
upper level energies around 1000\,K came as a surprise (see Fig.~1 and the Supplementary Figs.~1--5). These
high-excitation ortho-H$_2$O lines provide  a strong diagnostic
to understand the origin of water in carbon-rich envelopes.
Moreover, for the first time, para-H$_2$O lines have been detected from
the envelope of a carbon-rich AGB star (see Fig.~1 and the Supplementary Figs.~1--5).

The kinetic temperature in the envelope is typically around 2000\,K in the dense environment just above the stellar photosphere and decreases to $\sim$10\,K in the tenuous outer envelope (see  Fig.~2). The presence of high-excitation ortho-H$_2$O lines can only be
explained if water is present in the warm inner region of the
envelope, at radial distances closer than $\sim$15\,\Rstar\ (or $7.5 \times 10^{14}$\,cm for a stellar radius, \Rstar, of $5.1 \times 10^{13}$\,cm). This
immediately excludes the three mechanisms that only place water in
the intermediate or outer regions of the envelope as the source of the water origin (see Fig.~2). The mechanism involving the radiative association of O and H$_2$ in the outer envelope\cite{Agundez2006ApJ...650..374A} can also be ruled out in view of the low rate constant recently calculated for this
reaction\cite{2010CPL...485...56T}.  A possibility for the origin of water in the innermost regions\cite{GonzalesAlfonso2007ApJ...669..412G} of the envelope is pulsationally induced shocks which results in a chemical stratification different from thermodynamic equilibrium (TE) chemistry. However, IRC\,+10216 has a C/O abundance ratio of 1.4\emph{\cite{1994A&A...288..255W}}. Recent non-TE calculations\cite{Cherchneff2006A&A...456.1001C} have shown that for a carbon-rich star with an even lower C/O ratio of 1.1, water should be almost completely absent in the inner wind, and might only exist between 1 and 1.4\,\Rstar, with a [H$_2$O/H$_2$] peak abundance around $5 \times 10^{-5}$. Simulating this situation and assuming the (too high) water abundance of $5 \times 10^{-5}$ over the full region between 1 and 1.4\,\Rstar, yields water line predictions being a factor 3 to 10 too low compared to the PACS and SPIRE observations,  ruling out the shock-induced non-TE chemistry as possible cause of water.

\phantom{\cite{Duari1999AandA...341L..47D},\cite{McCabe1979Natur.281..263M},\cite{Glassgold1996ARA&A..34..241G}}

An alternative origin for  the warm water vapour could be provided by  
photochemistry in the inner regions of the
CSE of IRC\,+10216. For a strictly isotropic
and homogeneous mass loss process, the inner regions are well
protected against the interstellar ultraviolet (UV) radiation by the
circumstellar material located outwards (the visual extinction of
interstellar light is more than 100 magnitudes for the innermost regions in IRC\,+10216\cite{Agundez2006ApJ...650..374A}). Circumstellar envelopes are,
however, not perfectly spherical but have inhomogeneities and a
more or less clumpy structure. Observational evidence for the clumpy structure of the envelope around
IRC\,+10216 has been found both at small and large scales through observations at
near-infrared and visible wavelengths\cite{1998A&A...333L..51W,2006A&A...455..187L} as well as through
millimeter-wave observations of different molecules\cite{2003ApJ...582L..39F,2008ApJ...678..303D,1993A&A...280L..19G} (see Supplementary Information). The existence of a clumpy structure allows for a deeper penetration of
a fraction of interstellar UV photons into the inner circumstellar layers,
thus promoting dense and warm UV-illuminated
inner regions. In such an environment water can be formed from
the photodissociation of the major oxygen-carrier molecules,
mostly $^{13}$CO and SiO ($^{12}$CO is hard to photodissociate due
to self-shielding effects), and the subsequent liberation of
atomic oxygen, which then converts into water through
the chemical reactions:
\begin{equation}
 {\rm O + H_2 \rightarrow OH + H}\,,{\rm \ and}
\end{equation}
\begin{equation}
{\rm OH + H_2 \rightarrow H_2O + H}\,,
\end{equation}
which are only rapid enough at temperatures above $\sim$300 K. We
stress that this mechanism does not necessarily extend to the
whole inner CSE but only to those inner clumpy regions which
are more exposed to the interstellar UV field. For the case where 10\% of the total circumstellar mass is  illuminated by interstellar UV photons through a cone 
where matter only fills 70\% of the solid angle of arrival of interstellar light, we predict that water would be formed in the inner envelope with a
maximum abundance relative to H$_2$ in excess of 10$^{-7}$ (see Fig.~3 and the Supplementary Information). The predicted water line strengths for the case of a minor UV-illuminated component are shown in green in Fig.~1. The deduced amount of water is 0.003 earth masses.

AGB stars also have  a soft UV stellar radiation field\cite{Querci1974A&A....31..265Q,Gustafsson2008A&A...486..951G}. Pulsationally
induced shocks might generate a surplus of UV photons close to the stellar photosphere. However, even for a clumpy inner envelope, the densities just above the stellar photosphere are so high that the UV photons will be severely attenuated in the first few $10^{14}$\,cm,  in contrast to the outer envelope, where the material is much less dense.

The penetration of interstellar UV photons will yield the formation of hydrides, other than H$_2$O, in the inner envelope through successive
hydrogenation reactions of heavy atoms (nitrogen, carbon or sulphur). Ammonia (NH$_3$, see  Fig.~3) is an
interesting example as it has been observed in IRC +10216 and also
from the CSEs of oxygen-rich AGB stars with abundances relative to
H$_2$ in the range 10$^{-7}$$-$10$^{-6}$
$^($\emph{\cite{2006ApJ...637..791H,1995ApJ...448..416M}}$^)$,
 much larger than the $1\times10^{-12}$ predicted by thermochemical models\cite{1964AnTok...9.....T,Cherchneff1992ApJ...394..703C}. Other molecules, such as 
HC$_3$N, which are typically formed by photochemistry in the outer
layers, also have increased abundances in the inner regions, as seen in
Fig.~3. The predicted higher abundance of HC$_3$N in the inner regions is confirmed by our recent observations of HC$_3$N (see Fig.~4) with the IRAM 30\,m telescope at Pico Veleta, for which the line profiles point towards  the existence of a warm inner component. The discovery of high-excitation H$_2$O lines in the inner warm and dense envelope of an evolved carbon-rich star questions our knowledge of the envelope chemistry and outlines the importance of UV induced photochemistry in the CSEs of evolved stars. In the case of oxygen-rich environments, the same mechanism predicts high abundances of carbon-rich species, such as HCN, CH$_4$ and CS\cite{Agundez2010}, as has already been observed in several targets\cite{Bujarrabal1994A&A...285..247B,Decin2010a}.



\begin{addendum}
\item[Supplementary Information] is linked to the online version of the paper at www.nature.com/na\-ture.
\item[Author Contributions] L.D.\ identified the molecular lines, analyzed and interpreted the PACS and SPIRE data and performed the non-LTE (non-local thermodynamic equilibrium) radiative transfer computations; M.A.\ and J.C.\ were responsible for the chemical modelling; M.J.B.\ is main responsible for the Herschel MESS Key program observations performed with the SPIRE instrument; M.A.T.G.\ is PI of the Herschel MESS Key program, F.D.\ checked the non-LTE radiative transfer calculations, R.L.\ modelled the spectral energy distribution (SED); EDB identified molecular lines; P.R.\ and B.V.\ were responsible for the calibration of the PACS observations; R.W.\ and E.T.P.\ were responsible for the calibration of the SPIRE observations; the rest of the team members belong to the Herschel MESS consortium, the framework in which the Herschel PACS and SPIRE spectroscopic observations were performed.
\item[Author Information] Reprints and permissions information is available at www.nature.com/re\-prints.
 \item 
Herschel is an ESA space observatory with science instruments provided by European-led Principal Investigator consortia and with important participation from NASA. PACS has been developed by a consortium of institutes led by MPE (Germany) and including UVIE (Austria); KUL, CSL, IMEC (Belgium); CEA, OAMP (France); MPIA (Germany); IFSI, OAP/AOT, OAA/CAISMI, LENS, SISSA (Italy); IAC (Spain). This development has been supported by the funding agencies BMVIT (Austria), ESA-PRODEX (Belgium), CEA/CNES (France), DLR (Germany), ASI (Italy), and CICT/MCT (Spain). SPIRE has been developed by a consortium of institutes led by Cardiff Univ. (UK) and including Univ. Lethbridge (Canada); NAOC (China); CEA, LAM (France); IFSI, Univ. Padua (Italy); IAC (Spain); Stockholm Observatory (Sweden); Imperial College London, RAL, UCL-MSSL, UKATC, Univ. Sussex (UK); Caltech, JPL, NHSC, Univ. Colorado (USA). This development has been supported by national funding agencies: CSA (Canada); NAOC (China); CEA, CNES, CNRS (France); ASI (Italy); MCINN (Spain); SNSB (Sweden); STFC (UK); and NASA (USA). IRAM is supported by INSU/CNRS (France), MPG (Germany), and IGN (Spain).

 \item[Competing Interests] The authors declare that they have no competing financial interests.
 \item[Correspondence] Correspondence and requests for materials
should be addressed to Leen Decin (email: Leen.Decin@ster.kuleuven.be).
\end{addendum}

\newpage
\begin{figure}
 \begin{center}
\caption{\textbf{Unblended ortho- and para-water lines detected with Herschel in IRC\,+10216.}
39 ortho-H$_2$O and 22 para-H$_2$O lines are identified, including low and high-excitation lines (see also Supplementary Figs.~1--5).  Panel (a) shows the ground-state ortho-H$_2$O line as observed with SWAS\cite{Melnick2001Natur.412..160M} (units of intensity, $T_A^*$, in K versus velocity with respect to the local standard of rest, ${\rm{v_{LSR}}}$, being $-26$\,km\,s$^{-1}$ for IRC\,+10216). Panels (b) -- (i) show the continuum subtracted flux (in Jy) versus wavelength of observation for ortho-H$_2$O lines, while panels (l) --(o) show several para-H$_2$O lines.
The HCN $\nu_2=4$ contribution to the ortho-H$_2$O($4_{2,3}-3_{1,2}$)  line in panel (d)  is $\sim$15\,Jy.  \newline
The coloured lines show the non-local thermodynamic equilibrium predictions (see Supplementary Information) for the different chemical mechanisms proposed as cause of water vapour in the envelope of IRC\,+10216. 
For the red and orange lines, the predictions of ortho-H$_2$O  represent envelope models with a constant fractional abundance of ortho-H$_2$O (relative to H$_2$) out to $4\times10^{17}$cm, where it is photodissociated\cite{GonzalesAlfonso2007ApJ...669..412G}. The abundance of ortho-H$_2$O was derived from the SWAS observations (see panel (a)).
The red model simulates ortho-H$_2$O originating in the intermediate envelope, with inner envelope radius, R$_{\rm{int}}$, of $2.1 \times 10^{15}$cm and the derived  abundance, [ortho-H$_2$O/H$_2$], of $2.5 \times 10^{-7}$. This model applies both for the hypothesis of the vaporization of icy bodies and the Fischer-Tropsch catalysis mechanism, predicting water around a few times $10^{15}$cm. The orange model assumes ortho-H$_2$O to be generated  in the outer envelope, with R$_{\rm{int}}$ of $4.3 \times 10^{16}$cm and [ortho-H$_2$O/H$_2$] equal to $6.7 \times 10^{-7}$.  Finally, the green model shows the model predictions for ortho-H$_2$O present in the inner envelope, with a radial distribution  of the fractional abundance as shown in Fig.~3. The last row shows four para-water lines, where the blue line is based on the fractional abundance distribution as show in Fig.~3, using an ortho-to-para H$_2$O ratio of 3:1. } 
\end{center}
\end{figure}
{\clearpage}\newpage

\begin{figure}
\begin{center}
\caption{\textbf{Schematic overview (not to scale) of the
   envelope around a carbon-rich AGB star.} Several chemical processes are indicated at the typical temperature and radial distance from the star in the envelope where they occur. Shock-induced non-equilibrium chemistry takes
place in the inner wind envelope\cite{Duari1999AandA...341L..47D},
dust-gas  and ion-molecule reactions alter the abundances
in the intermediate wind zone, molecules such as CO and HCN
freeze-out at intermediate radii\cite{McCabe1979Natur.281..263M},
and the penetration of cosmic rays and ultraviolet (UV) photons
dissociates the molecules and initiates an active photochemistry
that creates radicals in the outer wind region\cite{Glassgold1996ARA&A..34..241G}. 
The different mechanisms hitherto proposed as origin of water in a carbon-rich environment are indicated at the bottom of the figure in grey at the typical distances where they occur. The penetration of the interstellar ultraviolet photons in a clumpy circumstellar environment is shown in blue. The resulting chemical processes important for the creation of H$_2$O and HC$_3$N are indicated in red.}
\end{center}
\end{figure}
{\clearpage}\newpage

\begin{figure}
\begin{center}
\caption{\textbf{Fractional abundances in the clumpy circumstellar environment of \irc.} The chemical model simulates a clumpy envelope structure, where a fraction of the interstellar UV photons is able to penetrate deep into the envelope (see Supplementary Information). This figure shows the predictions for the radial distribution of the fractional abundances relative to H$_2$ of H$_2$O, NH$_3$, and HC$_3$N for a model with a minor UV-illuminated component superposed on a major UV-shielded component. The minor component shown in this figure contains 10\% of the total circumstellar mass ($f_M=0.1$), which is illuminated by interstellar UV photons through a cone where matter fills 70\% of the solid angle of arrival of interstellar light ($f_{\Omega}=0.3$). Dashed lines correspond to the minor UV-illuminated component, continuous lines to the major UV-shielded component, and thick grey lines to the weighted average abundance over the two components. The weighted average abundance is computed as $\overline{X_i}(r)$ = ($1-f_M$)$X_i^{\rm major}(r)$ + ($f_M$)$X_i^{\rm minor}(r)$, where $X_i^{\rm major}(r)$ and
$X_i^{\rm minor}(r)$ are the abundance of the species $i$ in the
major UV-shielded and minor UV-illuminated component, respectively, as a function of radius
$r$. Note that for H$_2$O and NH$_3$  the contribution from the major UV-shielded component is negligible, and the dashed and thick grey line coincide.}
\end{center}
\end{figure}

\begin{figure}
\begin{center}
\caption{\textbf{HC$_3$N  as observed in the envelope of IRC\,+10216.}
The HC$_3$N J=29--28, J=33--32 and J=37-36 lines have been observed with the IRAM telescope. All three HC$_3$N lines show a clear flat-topped profile. The HC$_3$N J=29--28 line is shown in this figure, and its line profile is compared to that of the AlF J=8--7 line (black line). Both lines were observed with the same telescope, the same beam, the same pointing, and have been calibrated in the same way. While the AlF line profile is U-shaped, the HC$_3$N line is flat topped, clearly indicating that HC$_3$N arises from gas extending to inner radii. The grey lines show two model predictions: the dashed grey line corresponds to a model prediction only taking the major UV-shielded component into account, the full gray line shows the theoretical line profile for a model including both the major UV-shielded component and the minor UV-illuminated component. The spectrum is plotted in terms of intensity ($T_A^*$ in K) versus velocity with respect to the local standard of rest ($\rm{v_{LSR}}$ in km\,s$^{-1}$). The LSR stellar radial velocity of IRC\,+10216 is $-26$\,km\,s$^{-1}$.}
\end{center}
\end{figure}

{\clearpage}\newpage

\begin{figure}
\begin{center}
\includegraphics[width=183mm]{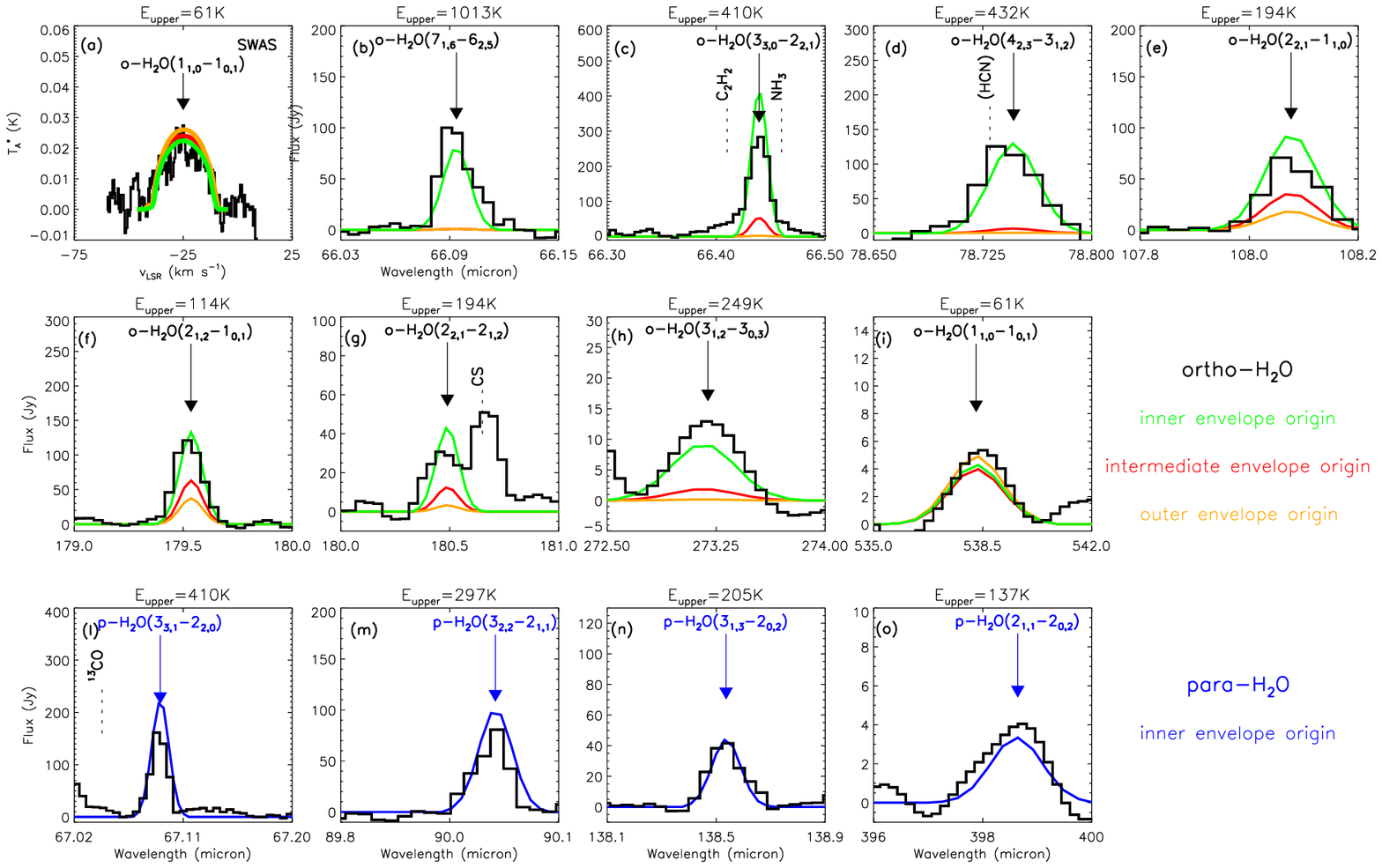}
\end{center}
\end{figure}

\clearpage
\newpage

\begin{figure}
\begin{center}
\includegraphics[width=12truecm]{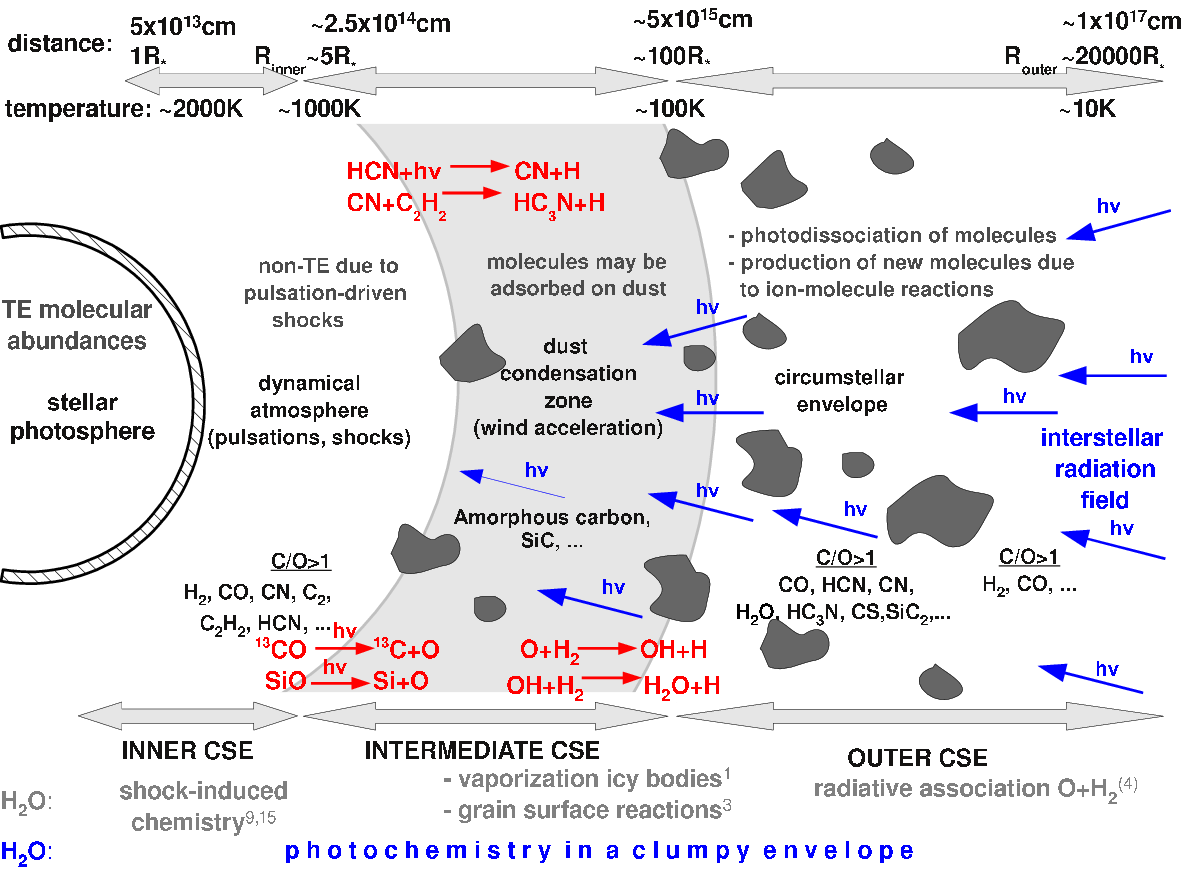}
\end{center}
\end{figure}

\clearpage
\newpage

\begin{figure}
\begin{center}
\includegraphics[angle=0,width=89mm]{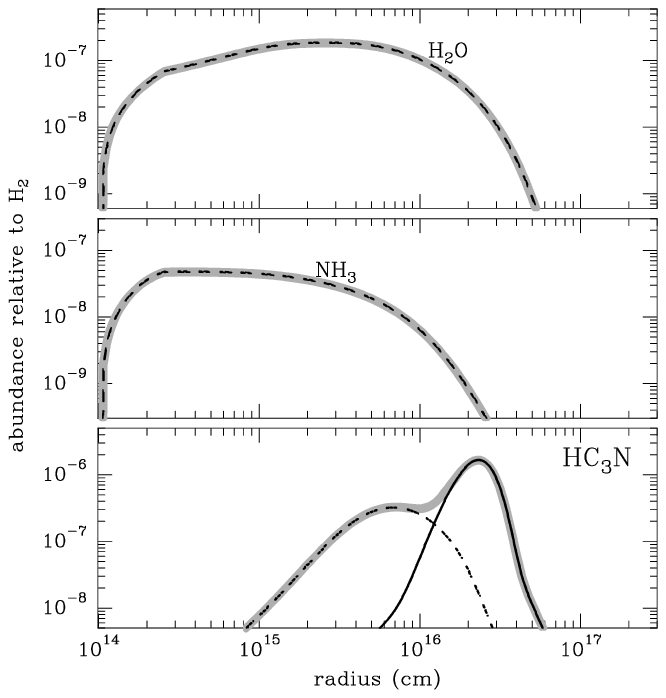}
\end{center}
\end{figure}

\clearpage
\newpage
\begin{figure}
 \begin{center}
\includegraphics[width=89mm]{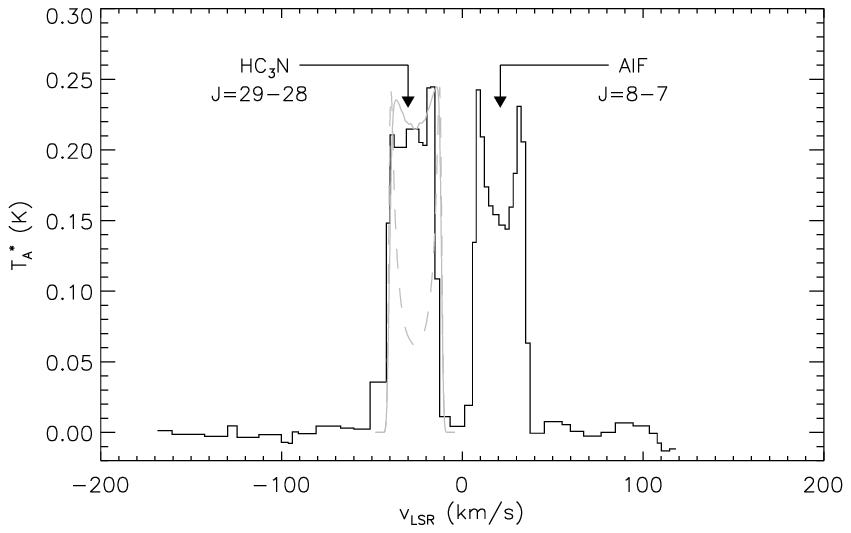}
\end{center}
\end{figure}

\end{document}


\maketitle 

\vspace*{-4cm}
\section*{1. Radiative transfer}
\vspace*{-1cm}
PACS and SPIRE observations  of IRC\,+10216 have been obtained in November 2009$^{13}$. The PACS data (55 -- 200\,$\mu$m) presented in the first paper describing the Herschel observations$^{13}$ were calibrated using a ground-based absolute flux calibration. Recently, a new absolute flux calibration for the PACS data has been obtained$^{12}$, yielding a decrease in absolute line flux of 30\% in the blue wavelength range ($\lambda < 98$\,$\mu$m) and of 10\% in the red wavelength range. The estimated calibration uncertainty on the PACS line fluxes is $\sim$30\%, but is higher for wavelengths shortward of 63\,$\mu$m due to non-optimal detector flatfielding. The PACS wavelength accuracy is around 20\%$^{11}$. For the SPIRE data (194 -- 670\,$\mu$m) the absolute flux calibration uncertainty is 15--20\% between 194 and 313\,$\mu$m,  and 20--30\% between 303 and 500\,$\mu$m, going up to 50\% beyond 500\,$\mu$m.

These newly calibrated data (see Supplementary Figs. 1--5) have been used to rederive the thermal structure of the envelope around IRC\,+10216 through non-local thermodynamic equilibrium (non-LTE) radiative transfer modelling of the $^{12}$CO lines, using the GASTRoNOoM code$^{29,}$\cite{Decin2006A&A...456..549D}. The PACS and SPIRE $^{12}$CO lines cover energy levels from $J=3$ (at 31 K) to $J=47$ (at 5853 K) and trace the envelope for radii $r\la1 \times 10^{17}$ cm ($r<2000$\,\Rstar).  The 60 lowest levels in the ground and first-vibrationally excited state are included in the radiative transfer modelling$^{29}$. The velocity structure was obtained by solving the momentum equation. The  (circum)stellar parameters are given in Table 1.  Specifically, we derive a gas mass-loss rate of $2.1 \times 10^{-5}$\,\Msun/yr for a [CO/H$_2$]-ratio of $1 \times 10^{-3}$$^($\emph{\cite{Zuckerman1986ApJ...304..394Z}}$^)$ and a kinetic temperature that is given by $T(r)=T_{\rm{eff}}(\Rstar) \times \left(r/\Rstar\right)^{-0.58}$ (for $r<10$\,\Rstar) and $T(r) \propto r^{-0.4}$ beyond that radius. The estimated uncertainty on the mass-loss rate is a factor of 2. The derived kinetic temperature agrees very well with the results obtained for the inner envelope from a detailed modelling of C$_2$H$_2$ and HCN\emph{\cite{Fonfria2008ApJ...673..445F}}. 

Using the derived thermodynamical structure, the level populations for ortho- and para-H$_2$O lines are calculated using the GASTRoNOoM-code for different stratifications of the water abundance throughout the envelope. The radiative transfer modelling includes the 45 lowest levels in the
ground state and first vibrational state (i.e.\ the bending mode $\nu_2\,=\,1$ at 6.3\,$\mu$m). Level energies, frequencies and
Einstein A coefficients are extracted from the Barber water line list\cite{Barber2006MNRAS.368.1087B}. The H$_2$O-H$_2$ collisional rates are taken from the Faure database\cite{Faure2007A&A...472.1029F}. The effect of including excitation to the first excited vibrational state of the asymmetric stretching mode ($\nu_3=1$) has been tested, and is found to be negligible. \\

\clearpage

\begin{center}
 \begin{figure}[!h]
\vspace*{3cm}
 \includegraphics{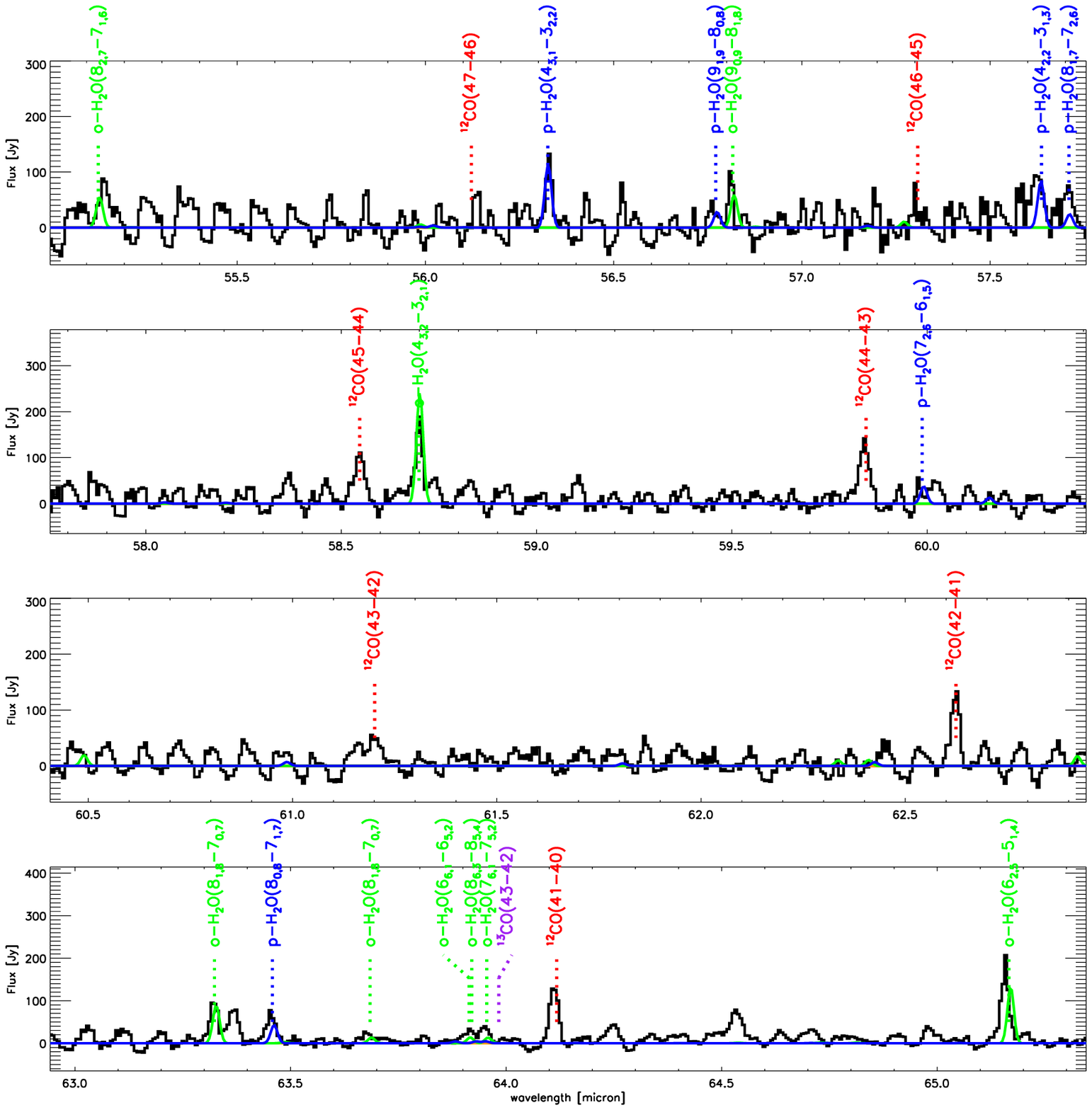}
\noindent Supplementary Figure 1: \textbf{Herschel spectrum of IRC\,+10216 compared to  water line predictions}. The Supplementary Figures 1--5 show the continuum subtracted Herschel PACS and SPIRE spectrum of IRC\,+10216 (in black).  The coloured lines are the model predictions for ortho- and para-water with parameters as described in the caption of Fig.~1 in the main journal. The detected water lines and the $^{12}$CO and $^{13}$CO lines are indicated. In total, 39 ortho-water lines and 22 para-water lines are detected in the PACS and SPIRE spectrum. Most of the water lines seen in the Supplementary Figures 1--5 are blended with another molecular emission line arising from HCN, SiO, SiS, CO, etc.; a sub-sample of (almost) unblended water lines is presented in Fig.~1 in the main journal.
\end{figure}
\end{center}

\begin{center}
 \begin{figure}[!htp]
\vspace*{3cm}
\includegraphics{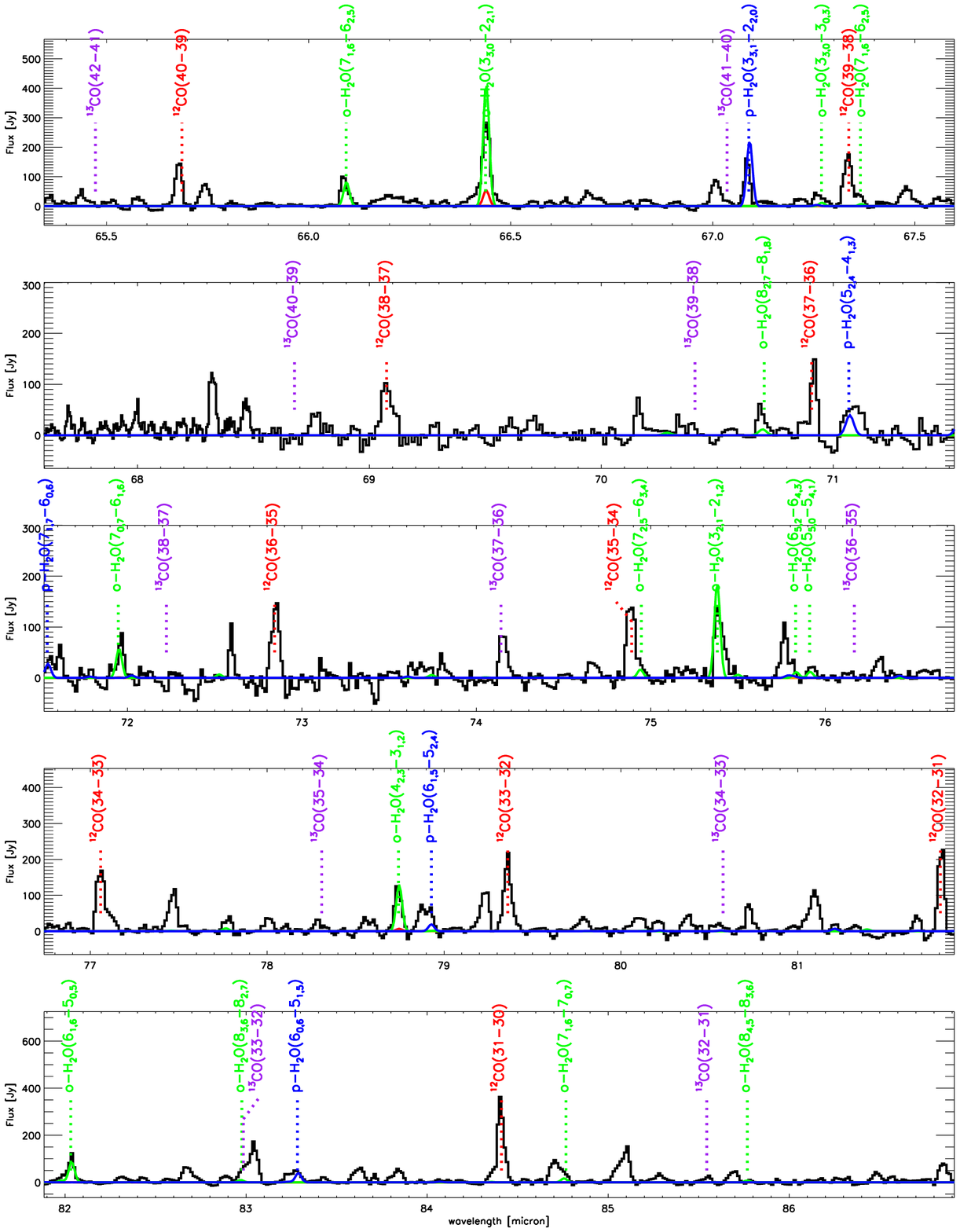}
\noindent Supplementary Figure 2: see caption of the Supplementary Figure 1. 
\end{figure}
\end{center}

\begin{center}
 \begin{figure}[!h]
\vspace*{3cm}
 \includegraphics{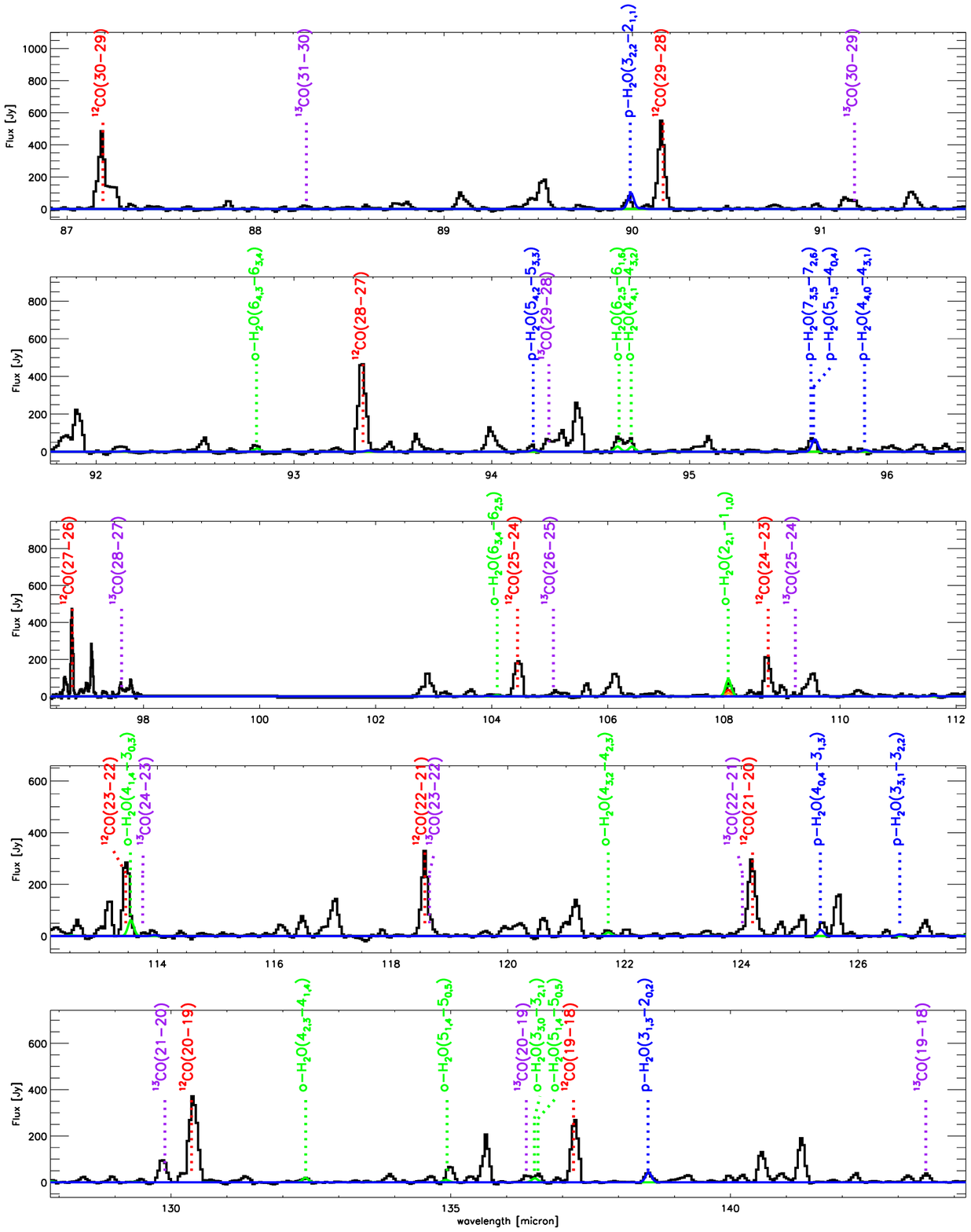}
\noindent Supplementary Figure 3: see caption of the Supplementary Figure 1. 
\end{figure}
\end{center}

\begin{center}
 \begin{figure}[!h]
 \vspace*{3cm}
\includegraphics{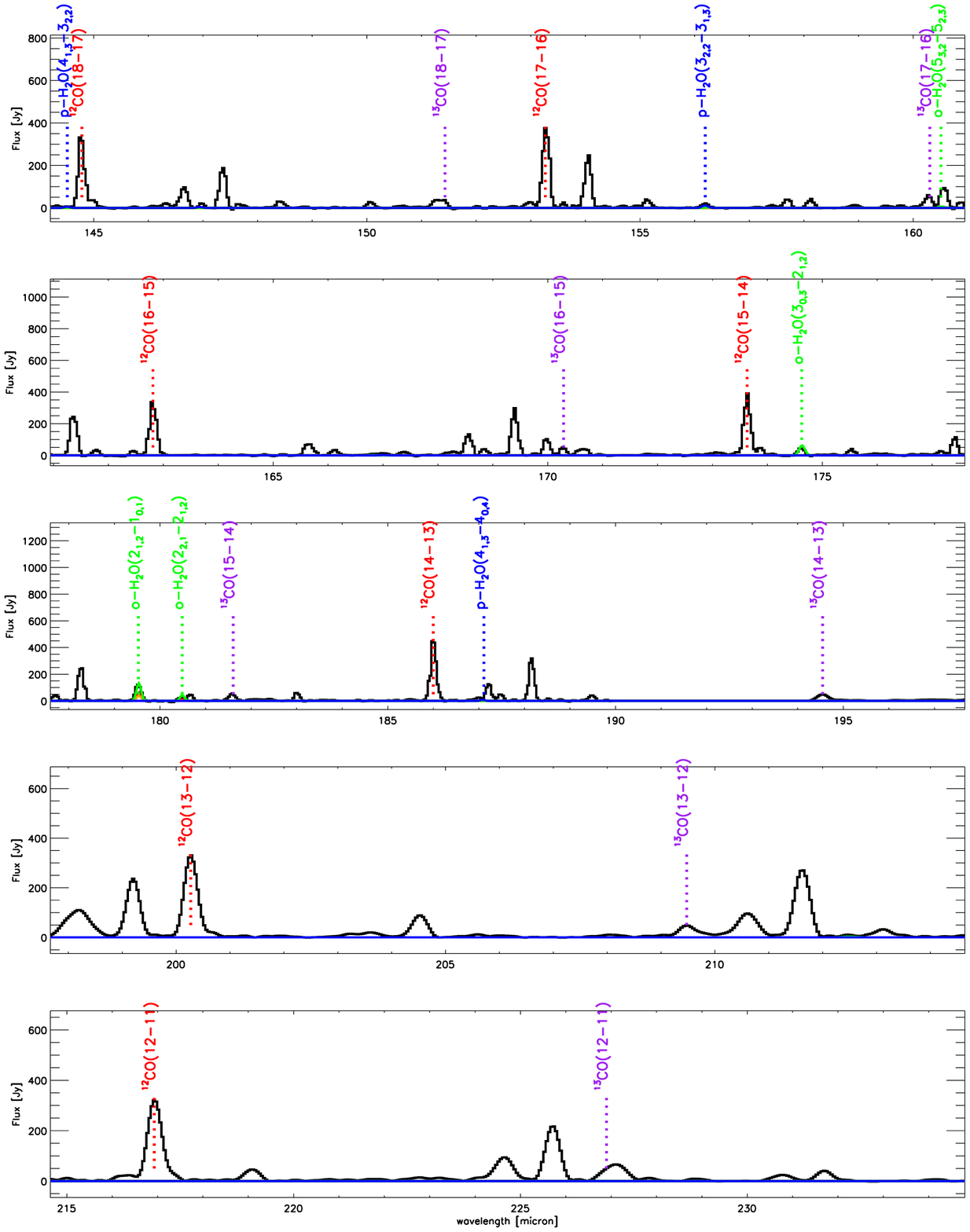}
\noindent Supplementary Figure 4: see caption of the Supplementary Figure 1. 
\end{figure}
\end{center}

\begin{center}
 \begin{figure}[!h]
 \vspace*{3cm}
\includegraphics{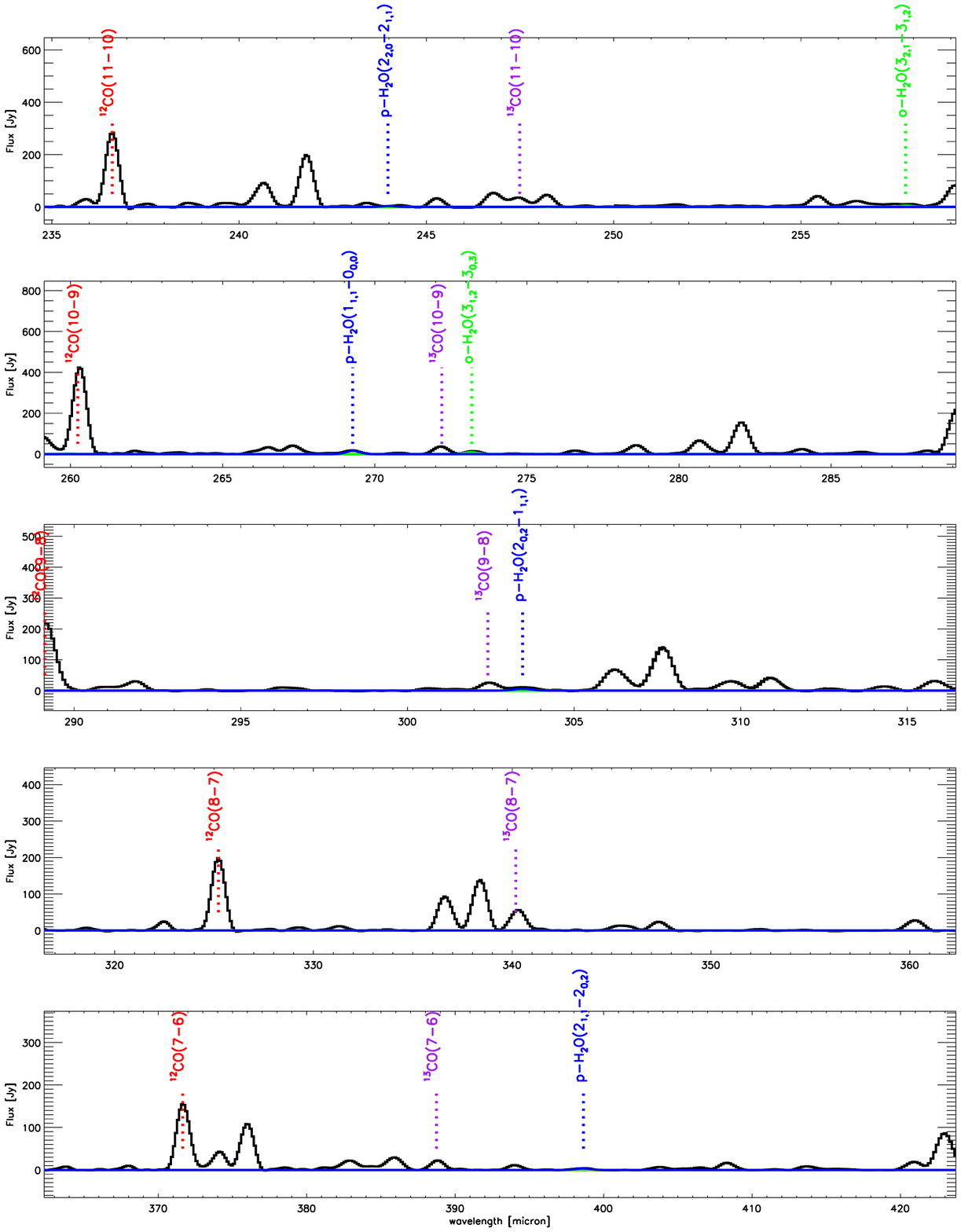}
\noindent Supplementary Figure 5: see caption of the Supplementary Figure 1. 
\end{figure}
\end{center}

\begin{center}
 \begin{figure}[!h]
 \vspace*{3cm}
\includegraphics{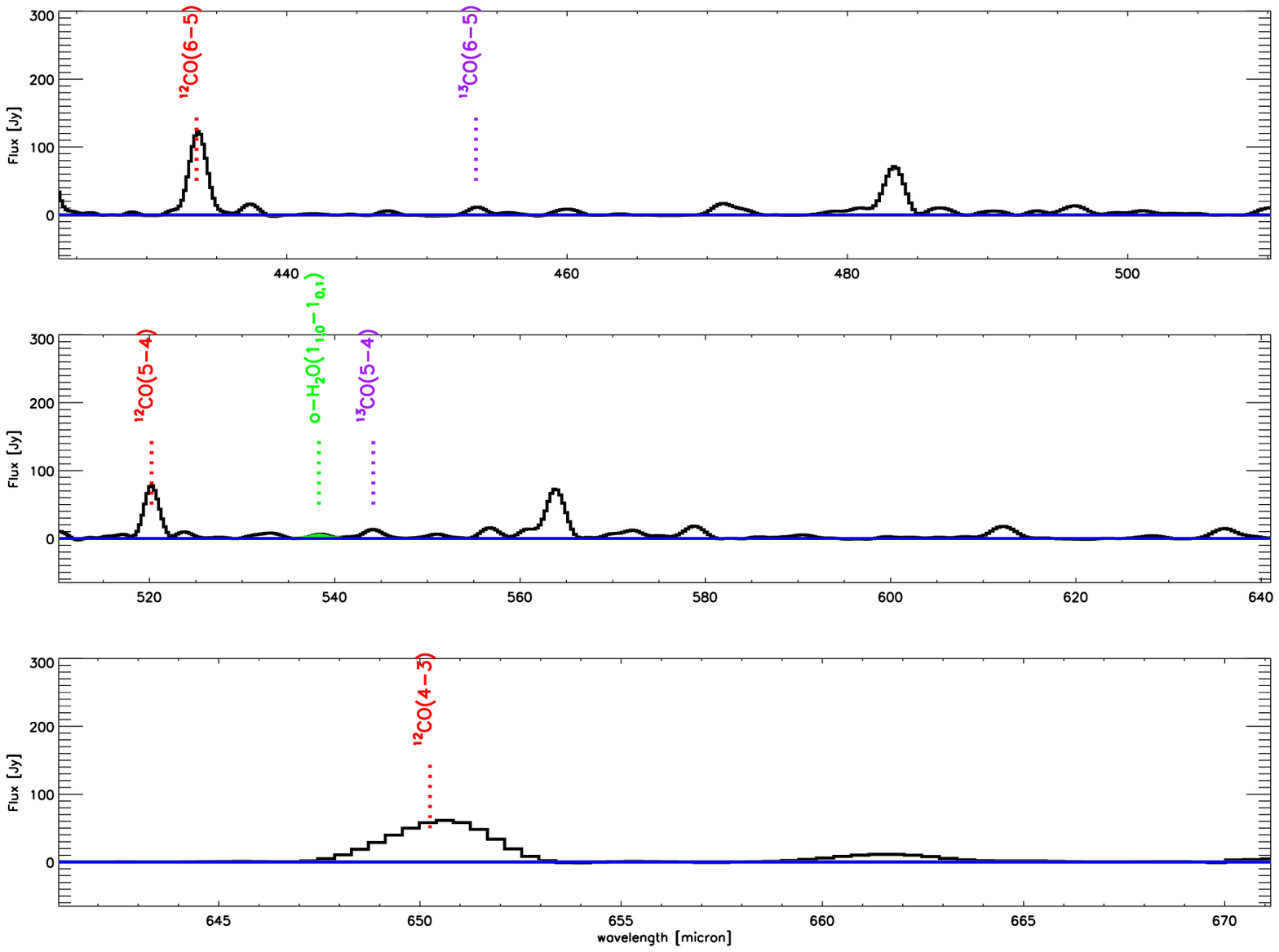}
\noindent Supplementary Figure 6: see caption of the Supplementary Figure 1. 
\end{figure}
\end{center}

\noindent Supplementary Table 1: \textbf{(Circum)stellar parameters of IRC\,+10216.}  \Teff\ is the effective stellar temperature, \Rstar\ denotes the stellar radius, $v_{\infty}$ the terminal velocity of the wind, \Mdot\ the gas mass-loss rate, and $R_{\rm{dust}}$ the dust condensation radius. 
\vspace*{-1ex}
\begin{table}[!h]
\setlength{\tabcolsep}{.8mm}
\label{Table:1}
\begin{center}
\begin{tabular}{lc|lc}
\hline \hline
\rule[0mm]{0mm}{5mm}\Teff [K] & {2330}$^($\emph{\cite{Ridgway1988ApJ...326..843R}}$^)$ & \Mdot [{\Msun}/yr] & $2.1 \times 10^{-5}$\\
$R_{\star}$ [$10^{13}$\,cm] & 5.1$^($\emph{\cite{1993ApJ...406..199K}}$^)$& $R_{\rm{dust}}$ [\Rstar] & {5.6}$^($\emph{\cite{Ridgway1988ApJ...326..843R}}$^)$  \\
$[$CO/H$_2]$ [$10^{-3}$]    & {1}$^($\emph{\cite{Zuckerman1986ApJ...304..394Z}}$^)$ &  $^{12}$CO/$^{13}$CO & 30$^{(13)}$ \\
distance  [pc] & 180$^($\emph{\cite{Fonfria2008ApJ...673..445F}}$^)$ &  $v_{\infty}$     [km\,s$^{-1}$]&  {14.5}$^($\emph{\cite{Debeck}}$^)$ \\ 
\hline
\end{tabular}
\end{center}
\end{table}

Continuum emission provides the dominant excitation source for H$_2$O, mainly through absorption of 6\,$\mu$m photons in the $\nu_2$ band and subsequent decay$^{7}$. The spectral energy distribution (SED) from 2\,$\mu$m to 1\,mm has been modelled using the MCMAX code\cite{Min2009A&A...497..155M}, a radiative transfer code specialised for the treatment of optically thick dusty circumstellar media. The relevant stellar and circumstellar input parameters are identical to the ones used in the molecular emission line modelling. The best fit to the ISO SWS and LWS (Infrared Space Observatory, Short Wavelength Spectrometer and Long Wavelength Spectrometer) spectrophotometric data between 2.4 and 197\,$\mu$m, supplemented with photometric data points\cite{lad2010}, requires the inclusion of amorphous carbon, iron, silicon carbide and magnesium sulfide as dust species. Assuming spherical grains, their respective derived  mass fraction abundances are 0.72, 0.05, 0.13 and 0.10. From these relative abundances, the mean specific density of the dust grains, the dust mass-loss rate and the wavelength-dependent absorption efficiencies are calculated to provide the dust component during the modelling of the molecular emission lines. The derived dust mass-loss rate is $8 \times 10^{-8}$\,\Msun/yr, yielding a dust-to-gas ratio of $\sim4\times10^{-3}$. Estimated uncertainties are a factor of 2. The mean specific dust density is 2.41\,g\,cm$^{-3}$.

The  water fractional abundance as derived from the SWAS observations (see Fig.~2) is slightly higher than results presented already in the literature$^{7}$. This mainly reflects a difference in (circum)stellar parameters and dust opacities. Using the same input parameters, similar abundances are derived.

\section*{2. Chemical modelling}

There is a lot of evidence for a clumpy structure of the envelope around IRC\,+10216 from astronomical observations carried out at small and large scales in
the visible and near-infrared wavelength ranges$^{17,18}$. These optical and near-infrared 
observations trace the dust density distribution, and thus the gas density distribution if both dust and gas are position coupled$^{20}$. Millimetre-wave interferometric observations have mapped the spatial distribution of  CO, finding a large scale clumpy
structure$^{19}$. Other molecules, such as CN, C$_2$H, C$_4$H, and HC$_5$N show a clumpy and hollow quasi-spherical distribution  with a lack of molecular emission in the NNE to SSW direction$^{20,21}$. The hollow distribution can be interpreted in terms of photochemistry of the parent species, creating reactive species which then undergo further chemical reactions\cite{Cherchneff1993ApJ...410..188C}. The conical holes in the NNE and SSW directions have been interpreted in terms of a molecular density minimum in that part of the envelope$^{12}$, or the fact that IRC\,+10216 is just leaving the AGB and that a bipolar outflow has dug out two symmetric conical holes at a position angle PA$\sim$20$^{\circ(}$\cite{Menshchikov2001A&A...368..497M}$^)$. The existence of these clumps and  conical holes opens a path for interstellar UV photons to penetrate deep into the envelope.

Whilst the existence of clumpy structures and cavities in the envelope of IRC\,+10216 is not in doubt according to the observations, it is difficult to model this complicated structure, since the different parameters characterizing the three-dimensional clumpy structure are not well constrained by the observations. 
We have thus adopted a simple modelling approach to simulate the clumpy structure and to investigate the effects on the circumstellar chemistry. We considered that the envelope consists of two different components: a major one which is well shielded against interstellar UV photons and only starts to undergo  photochemistry in the outer layers, and a minor one (which accounts for a fraction $f_M$ of the total circumstellar mass) for which the shielding material located in
the radial outward direction is grouped into clumpy structures leaving a fraction $f_\Omega$ of the solid angle of arrival of
interstellar light free of matter. We name these two components `the major UV-shielded' and `the minor UV-illuminated' component. As might be clear the minor UV-illuminated component is not completely exposed to UV photons, but is also partially shielded against UV photons by the clumps located on the way to the interstellar medium. Likewise, the major UV-shielded component starts to be partially exposed to UV photons in the outer envelope.

The physical conditions in both components are assumed to be the same, except that for the minor UV-illuminated component the visual extinction against interstellar light is reduced, as compared to that of the major UV-shielded component, due to the empty space between the clumpy material. The observations quoted in the previous paragraph have not provided an estimate of the density in the interclump medium. In our chemical model we have assumed that the density in the interclump medium is formally zero (in practice it is low enough to allow interstellar UV photons to penetrate without significant extinction). The clumpiness is assumed to extend down to the innermost regions of the envelope ($r\sim$10$^{14}$\,cm). 
Therefore, besides the standard physical parameters of any model describing a spherical circumstellar envelope (see Supplementary Table~1), we utilize two further phenomenological parameters to take care of the effects of clumpiness on the chemistry: for the minor UV-illuminated component $f_M$ is the fraction of the total circumstellar mass in the minor component and $f_\Omega$ is the fraction of the solid angle of arrival of interstellar light which is free of matter. 

The gas phase chemical composition of both the major UV-shielded and the minor UV-illuminated components are obtained from chemical kinetic calculations starting at the innermost layer ($r$\,=\,10$^{14}$\,cm) up to a distance of $r$\,=\,10$^{18}$\,cm, where most molecules have already been destroyed by interstellar UV photons. The chemical network utilized is essentially that used in previous work on the chemistry of oxygen in IRC\,+10216$^{4}$,
with updates for some reaction rate constants and photodissociation/ionization rates from the recent literature\cite{2006FaDi..133..231V,Woodall2007A&A...466.1197W}. The interstellar UV field has been decreased by a factor of 2 with
respect to the standard interstellar UV field\cite{Draine1978ApJS...36..595D}, to predict the angular
position of the maximum abundance of C$_2$H and CN at 15\arcsec\  and
20\arcsec\  from the star, respectively, in agreement with interferometric observations\cite{Agundez2006ApJ...650..374A}. We note that when fitting the peak
abundance distribution of these molecules, there exists a degeneracy between various physical parameters, such as the
interstellar UV field, the mass loss rate, the extinction law, and the distance to IRC\,+10216. Furthermore, the exact  value of
the local interstellar UV field around IRC\,+10216 is not precisely known. Recently, an extended UV structure surrounding IRC\,+10216
has been discovered at a radial distance of about 12\arcmin$^($\cite{Sahai2010ApJ...711L..53S}$^)$. This structure has been
attributed to the interaction of the stellar wind with the interstellar medium, and it might contribute to the ambient UV
field seen by the circumstellar envelope of IRC\,+10216. The exact value of the interstellar UV field around IRC\,+10216 is, however,
not critical in determining the flux of UV photons reaching the inner layers of the minor UV-illuminated component. What really counts is the degree of clumpiness, i.e.\ the parameter $f_\Omega$. A more detailed description of the chemical modelling approach for this kind of environment, with a major UV-shielded and a minor UV-illuminated component, for both carbon- and oxygen-rich circumstellar envelopes with different mass-loss rates will be presented in another article$^{27}$.

For the envelope around IRC\,+10216, the inner regions of the major UV-shielded component (surrounded by a smooth envelope) are indeed
well shielded against interstellar UV photons, with a visual extinction larger than 50\,mag for the regions down to 5\,\Rstar\cite{Agundez2006ApJ...650..374A}. 
Assuming that the minor UV-illuminated component is illuminated by interstellar UV photons in a cone where matter is just filling 50--80\% of the solid angle of arrival of interstellar light ($f_\Omega$ = 0.5--0.2), results in a substantial decrease of the effective visual extinction in the inner envelope from more than 50\,mag to less than 0.5\,mag, which indeed allows for photochemistry to take
place in the inner envelope. In the final model, whose results are shown in Fig.~3 of the main article, we have adopted a value of 0.3 for $f_\Omega$
and assumed that the minor UV illuminated component accounts for 
10\% of the total circumstellar mass ($f_M$ = 0.1). The resulting
abundance radial profile for each species $i$ are then obtained as
the weighted-average of the abundance in both components, i.e.\ as
$\overline{X_i}(r)$ = ($1-f_M$)$X_i^{\rm major}(r)$ +
($f_M$)$X_i^{\rm minor}(r)$, where $X_i^{\rm major}(r)$ and
$X_i^{\rm minor}(r)$ are the abundance of the species $i$ in the
major UV-shielded and minor UV-illuminated component, respectively, as a function of radius
$r$.  This approach assumes that the major UV-shielded and the minor UV-illuminated components are not
distributed over a given preferred radial direction but over many different radial directions, in such a way that we can average over all directions to get a mean value of the abundance $\overline{X_i}$ at each radius $r$. Whilst it is clear that in reality both components cannot coexist in the same radial
directions, this approach permits us to describe the distribution of the molecular abundances under spherical geometry, which facilitates the radiative transfer modeling described in Sect.~1. We note that the effect of a clumpy medium on spectral line formation can be characterized by a single number
multiplying the mean column opacity\cite{Conway2004}.  Even for an arbitrarily complex clump distribution, the effect can be reduced to a single scaling number. The scaling factor reduces the effective opacity of a clumpy medium compared to a smooth gas of the same mean column density. The scaling factor  is directly retrieved from the chemical modelling (see Fig.~3 in the main article) and taken into account when modelling the H$_2$O line profiles.

As shown in Fig.~3 of the main article, water vapour reaches a
weighted-averaged abundance relative to H$_2$ in excess of
10$^{-7}$. The photodissociation radius for each molecule is calculated from the chemical network taking the two components into account. According to the model water extends from the innermost regions ($r$ $\sim$ 10$^{14}$\,cm) up to a radial distance of (1--3) $\times$ 10$^{16}$\,cm. In this region water vapour keeps a
roughly constant abundance due to the balance between the formation processes (the reaction of O and H$_2$ followed by that
of OH and H$_2$) and the destruction processes (photodissociation by the ambient interstellar UV field). The photodissociation
radius of water, (1--3) $\times$ 10$^{16}$\,cm, is noticeably lower than that predicted by other hypotheses, 4 $\times$ 10$^{17}$\,cm
(see main article). The smaller photodissociation radius obtained by our model is fully consistent with the SWAS, PACS, and SPIRE observations, as shown in Fig.~2 of the main article.


\setcounter{enumiv}{30}
